\documentclass[12pt]{amsart}
\usepackage{amssymb}
\usepackage[mathscr]{eucal}
\newtheorem{prop}{Proposition}
\theoremstyle{definition}

\newcommand{\ds}{\displaystyle}

\def\EXP{\mbox{{\large\bf e}}}
\newcommand{\uop}{\mathbf{u}}
\newcommand{\wop}{\mathbf{w}}
\newcommand{\xop}{\mathbf{x}}
\newcommand{\zop}{\mathbf{z}}
\newcommand{\Xop}{\mathbf{X}}
\newcommand{\Yop}{\mathbf{Y}}
\newcommand{\Zop}{\mathbf{Z}}
\newcommand{\R}{\mathbf{R}}
\newcommand{\Q}{\mathbf{Q}}

\newcommand{\ltoda}{\ell}
\newcommand{\ldst}{{\widetilde{\ell}}}
\newcommand{\montoda}{\widehat{\mathbf{t}}}
\newcommand{\tp}{\mathbf{t}}
\newcommand{\Ltoda}{L}
\newcommand{\Ldst}{\widetilde{L}}
\newcommand{\Montoda}{\widehat{T}}
\newcommand{\Trtoda}{T}
\newcommand{\Tau}{\mbox{\Large $\tau$}}
\newcommand{\Kappa}{\mbox{\Large $\kappa$}}
\newcommand{\Mu}{\mbox{\Large $\mu$}}
\newcommand{\Om}{\Delta}

\newcommand{\omeg}{{\omega^{1/2}}}
\newcommand{\omegg}{{\omega^{-1/2}}}
\newcommand{\ff}{\phi}
\newcommand{\state}{\mathfrak{s}}
\newcommand{\smatrix}{\mathfrak{S}}

\begin{document}

\title[]{Relativistic Toda chain at root of unity}%
\author{S. Pakuliak \and S. Sergeev}%
\address{BLTP, JINR, Dubna 141980, Russia}%
\email{pakuliak@thsun1.jinr.ru}%
\email{sergeev@thsun1.jinr.ru}

\thanks{This work was supported by the grants RFBR N
01-01-00201, INTAS OPEN 97-01312, RFBR-CNRS PICS N 608/RFBR
98-01-22033 and RFBR N 00-01-00299}%

\begin{abstract}
We declare briefly several interesting features of the quantum
relativistic Toda chain at N-th root of unity. We consider the
finite dimensional representation of the Weyl algebra. The origin
of the features mentioned is that we consider simultaneously the
quantum finite dimensional part and the classical dynamics of
N-th powers of Weyl's elements. As the main result, using the
technique of Q-operators, we establish a correspondence between
the separation of variables in the quantum model and the
B\"acklund transformations of its classical counterpart.
\end{abstract}
\maketitle

\section*{Introduction}

The quantum relativistic Toda chain was the subject of paper
\cite{PS}. There it was considered for Weyl's parameter $q$
inside the unit circle. In this paper we collect several results
concerning this model with $q=\omega$ -- $N$-th primitive root of
unity, so that it is possible to realize Weyl's elements as
unitary $N\times N$ matrices. The quantum state space of the model
becomes the finite dimensional one. The relativistic Toda chain
at the root of unity is a relative of, say, chiral Potts model
\cite{CPM,BS} etc., being the most simple model in the hierarchy
of integrable models, associated with the cyclic representations
of the affine Lie algebras.

In this notes we collect several interesting features of the
relativistic Toda chain at the root of unity. The main result is
the proof of the possibility to construct explicitly the
eigenvectors of the off-diagonal element $\mathbf{b}(\lambda)$ of
the monodromy matrix, that is the starting point of the
separation of the variables method
\cite{Sklyanin-sep,Sklyanin-rev,KS-manybody,Lebedev}. The main
object in this construction is a kind of Baxter's $Q$-operator
\cite{Baxter,BS}. The method we use resembles the idea of
\cite{KS-manybody} to get a projector to an eigenstate of
$\mathbf{b}(\lambda)$ as a product of {\em operators} $Q$ with
special values of the spectral parameter. Actually we prove this
hypothesis, but it appear that in the projector one has to take
{\em modified} $Q$-operators which do not commute with the
transfer matrix and do not form the commutative family.

Usually at the root of unity one tries to fix the primitive
centers of a representation, working in the pure quantum space.
Our method implies the consideration of the classical dynamics of
these centers \cite{BR-unpublished}. This approach is natural
when one formulates a model in the terms of mappings, and then
makes the reduction $q\mapsto\omega$. The results of this note is
the application of more general scheme, derived for the three
dimensional models, to the most simple spatial structure, see
\cite{Sergeev}. In the more convenient language, we construct a
projector to an eigenstate of $\mathbf{b}(\lambda)$ in terms of
modified quantum $Q$-operators, whose classical counterparts
correspond to the B\"acklund transformations of the classical
model \cite{Sklyanin-rev}.

Mainly we do not prove the propositions and do not fall into the
details. The details are rather tedious, they are connected with
a lot of extra definitions and parameterizations. The aim of this
note is just to describe a general scheme of the investigations.
The details are the subjects of the forthcoming papers.

\section{Formulation of the model}

\subsection{Lax matrices}

Define the quantum Lax matrix of $m$-th site of the quantum
relativistic Toda chain as follows:
\begin{equation}\label{l-toda}
\ds\ltoda_m(\lambda)\;=\;\left(\begin{array}{rcr} \ds
1\,+\,{\kappa\over\lambda}\,\uop_m^{}\,\wop_m^{} &,&
\ds -\,{\omeg\over\lambda}\,\uop_m^{} \\
&&\\
\ds \wop_m^{} &,& 0 \end{array}\right)\;.
\end{equation}
Here $\lambda$ is the spectral parameter, and $\kappa$ is an
extra complex parameter, common for all sites. Elements $\uop_m$
and $\wop_m$ form the ultra-local Weyl algebra
\begin{equation}\label{Weyl-algebra}
\ds \uop_m^{}\;\wop_m^{}\;=\; \omega\;\wop_m^{}\;\uop_m^{}\;,
\end{equation}
and $\uop_m$, $\wop_m$ for different sites commute. Here Weyl's
parameter $\omega$ is the primitive root of unity,
\begin{equation}
\ds\omega\;=\;\EXP^{2\,\pi\,i/ N}\;,\;\;\;
\omeg\;=\;\EXP^{i\,\pi/N}\;.
\end{equation}
This means, the $N$-th powers of the Weyl elements are the
centers. We will deal with the finite dimensional unitary
representation of the Weyl algebra, i.e.
\begin{equation}\label{uw-XZ}
\ds \uop\;=\;u\;\xop\;,\;\;\;\wop\;=\;w\;\zop\;,
\end{equation}
where $u$ and $w$ are $\mathbb{C}$-numbers, and the popular
representation of $\xop$ and $\zop$ is e.g.
\begin{equation}\label{representation}
\ds \xop\;|\alpha\rangle\;=\;\omega^\alpha\;|\alpha\rangle
\;,\;\;\; \zop\;|\alpha\rangle\;=\;|\alpha+1\,\rangle\;,\;\;\;
\langle\alpha|\beta\rangle\;=\;\delta_{\alpha,\beta}\;.
\end{equation}
Thus $\xop$ and $\zop$ are normalized to the unity $N\times N$
dimensional matrices, and the $N$-th powers of the local Weyl
elements are $\mathbb{C}$-numbers
\begin{equation}\label{uw-N-powers}
\ds \uop_m^N\;=\;u_m^N\;\stackrel{def}{=}\;U_m\;,\;\;\;
\wop_m^N\;=\;w_m^N\;\stackrel{def}{=}\;W_m\;.
\end{equation}
In general, all $u_m$ and $w_m$ are different, so we deal with
the inhomogeneous chain. Index $m$ means the number of Weyl
algebra in the tensor product of $M$ copies, and the state of
this tensor product we denote as
\begin{equation}
\ds
|\alpha_1\rangle\otimes|\alpha_1\rangle\otimes\cdots\otimes|
\alpha_M\rangle\;
\stackrel{def}{=}\;|\alpha_1,\dots,\alpha_M\rangle\;\equiv\;
|\vec\alpha\rangle\;.
\end{equation}
The variables $U_m$ and $W_m$ form the classical counterpart of
the quantum relativistic Toda chain, and it is useful to consider
the classical Lax matrix
\begin{equation}\label{L-toda}
\ds\Ltoda_m\;\stackrel{def}{=}\; \left(\begin{array}{cc} \ds
1\,+\,{\Kappa\over\Lambda}\,U_m W_m & \ds
{U_m\over\Lambda}\\
&\\
\ds W_m & 0
\end{array}\right)\;.
\end{equation}
Here in the spirit of (\ref{uw-N-powers}) we have implied
\begin{equation}\label{kl-N}
\ds\Kappa\;\stackrel{def}{=}\;\kappa^N\;,\;\;\;\;
\Lambda\;\stackrel{def}{=}\;\lambda^N\;.
\end{equation}

\subsection{Integrability}

The quantum Lax matrices are to be multiplied in the natural
order,
\begin{equation}\label{l-monodromy}
\ds\montoda(\lambda)\;\stackrel{def}{=}\;
\ltoda_1(\lambda)\;\ltoda_2(\lambda)\;\cdots\;
\ltoda_M(\lambda)\;=\;
\left(\begin{array}{cc} \ds
\mathbf{a}(\lambda) & \ds
\mathbf{b}(\lambda)\\
&
\\ \ds \mathbf{c}(\lambda) &
\mathbf{d}(\lambda)\end{array}\right)\;,
\end{equation}
and the quantum transfer matrix for the periodic chain of the
length $M$ is the trace of the monodromy matrix
(\ref{l-monodromy}):
\begin{equation}\label{transfer}
\ds \tp(\lambda)\;=\;\mbox{tr}\;\montoda(\lambda)\;=\;
\sum_{k=0}^{M}\; \lambda^{-k}\;\tp_k\;.
\end{equation}
In particular, $\tp_0^{}=1$ and
$\tp_M=\kappa^M\prod_{m=1}^M\,\uop_m^{}\wop_m^{}$. The set of
$\tp_k$ is commutative, this is provided by the intertwining
relation for the Lax operators,
\begin{equation}\label{RLL}
\ds R(\lambda,\mu)\;\ltoda(\lambda)\otimes\ltoda(\mu) \;=\;
(1\otimes \ltoda(\mu))\;(\ltoda(\lambda)\otimes 1)\;R(\lambda,\mu)
\end{equation}
where it is implied the tensor product of two $2\times 2$
matrices with the identical Weyl algebra entries
$\uop,\wop,\kappa$, different spectral parameters $\lambda$ and
$\mu$, and
\begin{equation}
\ds R(\lambda,\mu)\;=\;\left(\begin{array}{cccc} \ds
\lambda\,-\,\omega\,\mu & 0 & 0 & 0 \\
0 & \ds \lambda\,-\,\mu & \ds \mu\,(1\,-\,\omega) & 0 \\
0 & \ds \lambda\,(1\,-\,\omega) & \ds
\omega\,(\lambda\,-\,\mu) & 0 \\
0 & 0 & 0 & \ds \lambda\,-\,\omega\,\mu
\end{array}\right)
\end{equation}
is the almost usual six-vertex $R$-matrix.

We have formulated the inhomogeneous chain in general, the
transfer matrix depends on $\lambda$, $\kappa$ and on set
$\{u_m,w_m\}_{m=1}^{M}$:
\begin{equation}
\ds\tp(\lambda)\;=\; \tp(\lambda,\kappa\,;\{u_m,w_m\}_{m=1}^M)\;.
\end{equation}
From (\ref{RLL}) it follows
\begin{equation}
\ds \Bigl[\;\tp(\lambda,\kappa\,;\{u_m,w_m\})\;,\;
\tp(\mu,\kappa\,;\{u_m,w_m\})\;\Bigr]\;=\;0\;.
\end{equation}

For the classical counterpart one may also define the monodromy
matrix
\begin{equation}\label{Mon-toda}
\ds\Montoda\;=\;\Ltoda_1\;\Ltoda_2\;\cdots\;\Ltoda_M\;=\;
\left(\begin{array}{cc} \ds A(\Lambda) & \ds B(\Lambda)\\
&\\
\ds C(\Lambda) & \ds D(\Lambda)
\end{array}\right)\;,
\end{equation}
and its trace
\begin{equation}
\ds \Trtoda(\Lambda)\;=\;A(\Lambda)\,+\,D(\Lambda)\;=\;
1+\sum_{k=1}^M \Lambda^{-k} T_M.
\end{equation}
Integrability of the classical model is provided by
\begin{equation}
\ds \{\,\Trtoda(\Lambda)\,,\,\Trtoda(\Lambda')\,\}\;=\;0\;,
\end{equation}
where $\{,\}$ means the Poisson braces, defined by
\begin{equation}\label{Poisson}
\ds \{U_n,W_m\}=\delta_{nm}U_n\,W_m\;.
\end{equation}
This Poisson bracket can be written in terms of $L$-operators
(\ref{L-toda})
\begin{equation}\label{Poisson-L}
\{L_n(\Lambda)\stackrel{\otimes}{,}L_m(\Lambda')\}=
\delta_{nm}[r(\Lambda,\Lambda'),L_n(\Lambda)\otimes L_m(\Lambda')]
\end{equation}
with classical trigonometric $r$-matrix
\begin{equation}
\ds r(\Lambda,\Lambda')\;=\;\frac{1}{2(\Lambda-\Lambda')}
\left(\begin{array}{cccc} \ds
\Lambda\,+\,\Lambda' & 0 & 0 & 0 \\
0 & \ds \Lambda\,-\,\Lambda' & \ds 2\Lambda & 0 \\
0 & 2\Lambda' & \ds \Lambda'\,-\,\Lambda & 0 \\
0 & 0 & 0 & \ds \Lambda\,+\,\Lambda'
\end{array}\right)
\end{equation}
Further in this letter we will never use the Hamilton
formalism  and the symbol $\{...\}$ will be reserved for the notion
of a set.

\subsection{Eigenvalues of $\mathbf{a}(\lambda)$,
$\mathbf{b}(\lambda)$, $\mathbf{c}(\lambda)$ and
$\mathbf{d}(\lambda)$.}

Admit now the following useful notation for the product over
$\mathbb{Z}_N$:
\begin{equation}
\ds \prod_{\lambda}\;f(\lambda)\;\stackrel{def}{=}\;
\prod_{n\in\,\mathbb{Z}_N}\;\;f(\omega^n\lambda)\;.
\end{equation}
The following proposition clarifies the relation between the
quantum model and its classical counterpart:
\begin{prop}\label{prop-abcd}
\begin{equation}
\begin{array}{l}
\ds \prod_{\lambda}\;\mathbf{a}(\lambda)\;=\;A(\Lambda)\;,\;\;\;
\prod_{\lambda}\;\mathbf{b}(\lambda)\;=\;B(\Lambda)\;,\\
\\
\ds \prod_{\lambda}\;\mathbf{c}(\lambda)\;=\;C(\Lambda)\;,\;\;\;
\prod_{\lambda}\;\mathbf{d}(\lambda)\;=\;D(\Lambda)\;.
\end{array}
\end{equation}
\end{prop}
Thus the spectra of the quantum operators $\mathbf{a}(\lambda)$ ...
$\mathbf{d}(\lambda)$ can be expressed through the classical
objects $A(\Lambda)$ ... $D(\Lambda)$.

\subsection{Gauge fixing}

For the subsequent considerations we need to introduce a couple
of notations and to fix some gauge degree of freedom. It is
useful to consider following three operators for the whole spin
chain:
\begin{equation}\label{XYZ}
\ds \Xop\,=\,\prod_{m=1}^{M}\,\xop_m^{}\;,\;\;\;
\Zop\,=\,\prod_{m=1}^{M}\,\zop_m^{}\;,\;\;\;
\Yop\,=\,\prod_{m=1}^{M}\,(-\omegg\,\xop_m^{}\zop_m^{})\;.
\end{equation}
$\Xop$ and $\Zop$ obey the following relations:
\begin{equation}\label{t-XZ}
\ds \tp(\lambda)\,\Xop\;=\;\Xop\,\tp(\omega\lambda)\;,\;\;\;
\tp(\lambda)\,\Zop\;=\;\Zop\,\tp(\omega^{-1}\lambda)\;.
\end{equation}
One can  see that the corresponding to $\Trtoda_M$ flow results
to the trivial re-scaling of the classical amplitudes $U_m$ and
$W_m$: $U_m\mapsto CU_m$, $W_m\mapsto C^{-1}W_m$. Since
$\tp_M^N=(-)^{M(N-1)}\Trtoda_M$ we may always re-define $\lambda$
so that
\begin{equation}\label{det-normal}
\ds \prod_{m=1}^{M}\;\omeg\,u_m^{}\,w_m^{}\;\equiv\;1\;,\;\;\;
\prod_{m=1}^M\;(-U_mW_m)\;=\;1\;.
\end{equation}
Also the gauge may be fixed in the most useful way. Define the
homogeneous chain via ($\epsilon=(-)^N$)
\begin{equation}\label{uw-homo}
\ds u_m\;=\;-\omega^{-1/2}\;,\;\;w_m=-1\;,\;\;
U_m=-\epsilon\;,\;\; W_m\;=\;{1\over\epsilon}\;,
\end{equation}
whereas the inhomogeneous chain's $u_m$ and $w_m$ we parameterize
via sets of $\tau_m$ and $\theta_m$:
\begin{equation}\label{uw-inhomo}
\ds u_m\;=\;-\omega^{-1/2}\,{\tau^{}_{m-1}\over\tau^{}_m}\;,
\;\;\; w_m\;=\;-\,{\theta^{}_m\over\theta^{}_{m-1}}\;,\;\;\;
m\,\in\,\mathbb{Z}_M\;.
\end{equation}

\subsection{Spectral curves}
Consider the spectral curve for the classical part of our model
By definition, and taking into account (\ref{det-normal}),
\begin{equation}\label{J-spectral}
\ds J(\Lambda,\Mu)\;\stackrel{def}{=}\;
-\Mu\det(\Mu^{-1}-\Montoda(\Lambda))\;=\; \Trtoda(\Lambda)
\,-\,{1\over\Mu}\,-\, {\Mu\over\Lambda^M}\;.
\end{equation}
Condition $J=0$ in the terms of $\Lambda,\Mu$ defines the genus
$g_M=M-1$ hyperelliptic classical spectral curve.

Our interest is the quantum model, and the quantum curve
$\Gamma_M$ is to be defined as
\begin{equation}\label{M-curve}
\ds (\lambda,\mu)\,\in\,\Gamma_{M}\;\;\;
\Leftrightarrow\;\;J(\lambda^N,\mu^N)\;=\;0\;.
\end{equation}
Quantum curve is not hyperelliptic, and its genus in general
position is $g_{N,M}=N^2M-2N+1$. The quantum curve will appear in
Baxter's equation.

\subsection{Homogeneous chain}

Now consider the homogeneous quantum and classical chains. The
values of $u_m,u_m$ are given by (\ref{uw-homo}). Due to
proposition (\ref{prop-abcd}), the spectrum of
$\mathbf{b}(\lambda)$ is defined by the value of $B(\Lambda)$.
We can calculate
\begin{equation}\label{B-spectrum}
\ds
B(\Lambda)\;=\;-\,\epsilon\,{1\over\Lambda}\,\prod_{k=1}^{M-1}\,
\left(1\,-\,{\Lambda(\phi_k)\over\Lambda}\right)\;,
\end{equation}
where
\begin{equation}\label{phi-k}
\ds\phi_k\;\; \in\;\; \{{\,\pi\over M}\,,\;{2\pi\over
M}\,,\;...\;,\;{(M-1)\pi\over M}\,\}\;,
\end{equation}
and the function $\Lambda(\phi)$ is defined by
\begin{equation}\label{phi-parametrization}
\ds\left\{\begin{array}{l} \ds
\Om(\phi)\;=\;\EXP^{i\,\phi}\,\left(
\sqrt{\cos^2\phi\,+\,\Kappa}\;+\;\cos\phi\right)\;,\\
\\
\Om^*(\phi)\;=\;\EXP^{-i\,\phi}\,\left(
\sqrt{\cos^2\phi\,+\,\Kappa}\;+\;\cos\phi\right)\;,\\
\\
\ds \Lambda(\phi)\;=\;\Om(\phi)\;\Om^*(\phi)\;.
\end{array}\right.
\end{equation}
From (\ref{B-spectrum}) it follows that up to inessential
$N$-dimensional block the spectrum of $\mathbf{b}(\lambda)$ is
described by the set of $M-1$ phases of
\begin{equation}\label{lambda-phi-k}
\ds\lambda_{\phi_k}\;:\;\lambda_{\phi_k}^N\;=\;\Lambda(\phi_k)\;.
\end{equation}
The reason for the above mentioned block which correspond
to the spectrum of $\xop_M$ is the same as in the usual
quantum Toda chain.

The curve $\Gamma_{M}$ now factorizes into $\Gamma_1^M$:
\begin{equation}
\ds J(\Lambda,\Mu)\;=\;-\;{1\over\Mu\Lambda^M}\;
\prod_{\displaystyle{\Om^M}=\Mu}\; (\Om^2-\Om(\Lambda-\Kappa)+\Lambda)\;,
\end{equation}
where the product is taken over all $M$-th roots of unity. In
particular, $J(\Lambda,\Mu)=0$ if there exists $\phi$:
$\Lambda=\Lambda(\phi)$ and $\Mu\;=\;\Om(\phi)^M$.

\section{Isospectrality problem}

This paper is devoted actually to a special class of
inhomogeneous chains. In this section we will give some notations
and formulate a proposition, which may be proved by the methods,
described in the subsequent sections.

At first, define ``solitonic tau - functions'' recursively as
\begin{equation}
\ds\begin{array}{l} \ds \Tau_m^{(0)}\;=\;1\;,\\
\\
\ds \Tau_m^{(1)}\;=\;1\,+\,f_1\,\EXP^{2 i m \phi_1}\;,\\
\\
\ds \Tau_m^{(2)}\;=\; 1+f_1\EXP^{2 i m \phi_1}+ f_2\EXP^{2 i m
\phi_2}+ d_{1,2} f_1 f_2 \EXP^{2 i m (\phi_1+\phi_2)}\;,
\end{array}\end{equation}
and so on,
\begin{equation}\label{Tau}
\ds \Tau_m^{(n)}\;=\;\Tau_m^{(n-1)}(\{f_k,\phi_k\}_{k=1}^{n-1})+
f_n\EXP^{2 i m
\phi_n}\Tau_m^{(n-1)}(\{d_{k,n}f_k,\phi_k\}_{k=1}^{n-1})\;,
\end{equation}
where all $\phi_k$ are different and belong to the set
(\ref{phi-k}). The phase shift $d_{j,k}=d(\phi_j,\phi_k)$ is
given by (see (\ref{phi-parametrization}))
\begin{equation}\label{phase-shift}
\ds d(\phi,\phi')\;=\; {\left(\Om(\phi)-\Om(\phi')\right)\;
\left(\Om^*(\phi)-\Om^*(\phi')\right) \over
\left(\Om(\phi)-\Om^*(\phi')\right)\;
\left(\Om^*(\phi)-\Om(\phi')\right) }\;.
\end{equation}
Complex number $f_k$ we will call the amplitude of $k$-th solitonic wave.
Set (\ref{phi-k}) is finite, so the number of solitons can't exceed
its maximal value $M-1$. The complete
$(M-1)$-solitonic tau-function is
\begin{equation}
\ds \Tau_m^{}\;\stackrel{def}{=}\;
\Tau_m^{(M-1)}\;=\;\Tau_m(\{f_k\}_{k=1}^{M-1})\;.
\end{equation}
Obviously, the less then $(M-1)$-solitonic tau-functions
are just particular cases of the complete one,
which can be obtained by the vanishing the corresponding
amplitudes.

Define 
a couple of useful functions:
\begin{equation}\label{c-function}
\ds c(\phi)\;=\; {\Om^*(\phi)\over\Om(\phi)}\;
{1\,-\,\Om(\phi)\over 1\,-\,\Om^*(\phi)}\;,
\end{equation}
and
\begin{equation}\label{s-function}
\ds
s(\phi,\phi')\;=\;{\Om^*(\phi)\over\Om(\phi)}\;{\Om^*(\phi')
\,-\,\Om(\phi)\over
\Om^*(\phi')\,-\,\Om^*(\phi)}\;.
\end{equation}
For given $\{f_k\}_{k=1}^{M-1}$ besides $\Tau_m$ we need
\begin{equation}\label{Theta}
\ds \Theta_m^{}(\{f_k\}_{k=1}^{M-1})\;\stackrel{def}{=}\;
\Tau_m(\{c_k f_k\}_{k=1}^{M-1})\;,
\end{equation}
where $c_k=c(\phi_k)$.

Consider now the inhomogeneous relativistic Toda chain, whose
parameters are given by (\ref{uw-inhomo}) with
\begin{equation}
\ds \tau_m^N\;=\;\Tau_m^{}(\{f_k\}_{k=1}^{M-1})\;,\;\;\;
\theta_m^N\;=\;\Theta_m^{}(\{f_k\}_{k=1}^{M-1})
\end{equation}
with arbitrary set of $\{f_k\in\mathbb{C}\}_{k=1}^{M-1}$. For the
sake of shortness we shall use the symbolic notation $\state$,
implying the definition of the solitonic data
$\{f_k\}_{k=1}^{M-1}$, construction of $\Tau_m$ and $\Theta_m$ via
eqs. (\ref{Tau},\ref{Theta}), choice of the phases for $\tau_m$
and $\theta_m$, and finally -- the definition of $u_m,w_m$, eq.
(\ref{uw-inhomo}):
\begin{equation}\label{STATE}
\ds \state\;:\;\biggl(\{f_k\}_{k=1}^{M-1}\mapsto\;
\{\Tau_m,\Theta_m\}\;\mapsto\; \{\tau_m,\theta_m\}\;
\mapsto\{u_m,w_m\}_{m=1}^M\biggr)\;.
\end{equation}
We will use the symbol $\state$ instead of sets $\{f_k\}$,
$\{\tau_m,\theta_m\}$ and so on. The quantum transfer matrix we
will denote as $\ds\tp(\lambda,\kappa;\state)$. In part, when all
$f_k=0$, this $\tp$ becomes the transfer matrix for the
homogeneous chain. The state, corresponding to the homogeneous
chain, we will denote as $\state_0$.

\begin{prop}\label{prop-isospectral}
For the values of $\{f_k\}$ being in general position, the
eigenvalues of $\tp(\lambda,\kappa;\state)$ do not depend neither
on $\{f_k\}$ nor on the phases of $\tau_m,\theta_m$. In other
words, for two states $\state$ and $\state'$ there exists a
matrix $\underset{\state\mapsto\state'}{\smatrix}$, invertible in
general position, such that
\begin{equation}
\ds\tp(\lambda,\kappa;\state)\;
\underset{\state\mapsto\state'}{\smatrix}\;=\;
\underset{\state\mapsto\state'}{\smatrix}\;
\tp(\lambda,\kappa;\state')
\end{equation}
\end{prop}
Actually the proof of this proposition is based on the analogous
proposition for the classical counterpart $\Trtoda(\Lambda)$. We
have solved the problem of the isospectrality for the homogeneous
chain, i.e. for the rational classical spectral curve. If one
starts from the classical spectral curve in general position, then
isospectral $\Tau_m$ and $\Theta_m$ would become theta-functions
on the jacobian of the spectral curve. Actually our ``solitons''
are the rational limit of the finite gap solution.

\section{Auxiliary Lax operator}

In this section we give a scheme for the construction of
operators $\smatrix$, intertwining the states $\state$ and
$\state'$.

\subsection{``Dimer Self Trapping'' Lax matrix}

Define the quantum auxiliary lax matrix, acting in the space
$\ff$ as follows:
\begin{equation}\label{ldst-uwkappa}
\ds\begin{array}{l} \ds
\ldst^{}_\ff(\lambda,\lambda_\ff)\;=\\
\\
\ds =\;\left(\begin{array}{ccc} \ds
1\,-\,\omeg\kappa_\ff^{}{\lambda_\ff^{}\over\lambda} \wop_\ff^{}
&,& \ds
-{\omeg\over\lambda}(1-\omeg\kappa_\ff^{}\wop_\ff^{})\uop_\ff^{}
\\
&&\\
\ds -\omeg\lambda_\ff^{}\uop_\ff^{-1}\wop_\ff^{} &,& \ds
\wop_\ff^{}
\end{array}\right)\;.\end{array}
\end{equation}
Here as previously
\begin{equation}
\ds\uop_{\ff}\;=\;u_\ff\;\xop_\ff\;,\;\;\;
\wop_\ff\;=\;w_\ff\;\zop_\ff\;.
\end{equation}
Also, as previously, define the classical counterpart $\Ldst$ of
$\ldst$. $U_\ff^{}=\uop_\ff^N$, $W_\ff^{}=\wop_\ff^N$,
$\Lambda_\ff^{}=\lambda_\ff^N$ etc:
\begin{equation}\label{L-dst}
\ds \Ldst_\ff\;=\;\left(\begin{array}{ccc} \ds
1+\Kappa_\ff^{}{\Lambda_\ff\over\Lambda}W_\ff^{} &,&
\ds{U_\ff\over\Lambda}(1+\Kappa_\ff W_\ff)\\
&&\\
\ds \Lambda_\ff{W_\ff\over U_\ff} &,& W_\ff
\end{array}\right)\;.
\end{equation}

\subsection{Quantum Darboux transformation}

Darboux transformation for the classical chain is the following
relation:
\begin{equation}\label{LL-darboux}
\ds\Ldst(U_{\ff,m}^{},W_{\ff,m}^{}) \Ltoda(U_m^{},W_m^{})\;=\;
\Ltoda(U_m',W_m') \Ldst(U_{\ff,m+1}^{},W_{\ff,m+1}^{})\;,
\end{equation}
where all contents of $\Lambda$, $\Lambda_\ff$, $\Kappa$,
$\Kappa_\ff$ are implied. Recall, (\ref{LL-darboux}) must be
identity with respect to $\Lambda$, while $\Lambda_\ff$ enters
into the mapping
\begin{equation}\label{darboux-mapping}
\ds u_m,w_m,u_{\ff,m},w_{\ff,m}\;\;\mapsto\;\;
u_m',w_m',u_{\ff,m+1},w_{\ff,m+1}\;.
\end{equation}
The quantum Darboux transformation is the following intertwining
relation:
\begin{equation}\label{op-recursion}
\ds\begin{array}{l}
\ds\ldst_\ff^{}(u_{\ff,m}^{},w_{\ff,m}^{})\cdot
\ltoda_m^{}(u_m^{},w_m^{})\; \R_{m,\ff}^{}\;=\\
\\
\ds =\; \R_{m,\ff}^{}\;\ltoda_m^{}(u_m',w_m')\cdot
\ldst_\ff^{}(u_{\ff,m+1}^{},w_{\ff,m+1}^{})\;.
\end{array}
\end{equation}
\begin{prop}
Eq. (\ref{LL-darboux}) is the admissibility condition for eq.
(\ref{op-recursion}). In other words, if mapping
(\ref{darboux-mapping}) solves (\ref{LL-darboux}), then there
exists unique $N^2\times N^2$ matrix $\R_{m,\ff}$, solving eq.
(\ref{op-recursion}).
\end{prop}
We do not give the explicit form of the matrix elements of
$\R_{m,\ff}$ here, it is not necessary in this brief notes.

Eq. (\ref{op-recursion}) is to be iterated for the whole chain. It
arises the monodromy of $\ltoda_m$ and the monodromy of
$\R_{m,\ff}$,
\begin{equation}\label{R-monodromy}
\ds \widehat{\Q}_\ff\;\stackrel{def}{=}\;
\R_{1,\ff}\,\R_{2,\ff}\,\dots\,\R_{M,\ff}\;,
\end{equation}
The monodromies obey the following relation:
\begin{equation}\label{ltQ}
\ds\begin{array}{l}
\ds\ldst_\ff^{}(\lambda;u_{\ff,1},w_{\ff,1})
\cdot\montoda(\lambda;\{u_m,w_m\})\;
\widehat{\Q}_\ff^{}\;=\\
\\
\ds \widehat{\Q}_\ff^{}\; \montoda(\lambda;\{u_m',w_m'\})\cdot
\ldst_\ff^{}(\lambda;u_{\ff,M+1},w_{\ff,M+1})\;.
\end{array}
\end{equation}
Cyclic boundary conditions imply
\begin{equation}\label{cyclic-recursion}
\ds u_{\ff,M+1}\;=\;u_{\ff,1}\;,\;\;\;w_{\ff,M+1}\;=\;w_{\ff,1}\;.
\end{equation}
In general (\ref{cyclic-recursion}) have many solutions (they are
subject of the next proposition). Suppose we have chosen an
appropriate branch. The trace of the monodromy (\ref{R-monodromy})
\begin{equation}\label{R-trace}
\ds \Q_\ff\;=\;\mbox{tr}_\ff\,\widehat{\Q}_\ff
\end{equation}
remembers the branch mentioned, and the reminder is the subscribe
`$\ff$' of $\Q_\ff$. Moreover we will regard $\phi$ as the
spectral parameter of $\Q_\ff$, implying
$\lambda_\ff^N=\Lambda(\phi)$, see. eq.
(\ref{phi-parametrization}), etc. Also one may show, parameter
$\Kappa_\ff$ is responsible for a gauge transformation for
$u_m',w_m'$. This may be neglected by the choice
$\Kappa_\ff=\Kappa/\Om_\ff$. Matrix elements of $\Q_\ff$ may be
written out explicitly, and their parameterization implies all
the information about $\lambda_\ff$ and $\{u_m,w_m\}$ and
$\{u_m',w_m'\}$ etc. Equation
(\ref{ltQ}) with the cyclic boundary conditions
(\ref{cyclic-recursion}) provides
\begin{equation}\label{t-Q-permutation}
\ds \tp(\lambda\,;\{u_m^{},w_m^{}\})\,\cdot\, \Q_\ff\;=\;
\Q_\ff\,\cdot\, \tp(\lambda\,;\{u_m',w_m'\})\;.
\end{equation}
It is useful to write this relation symbolically in the form
\begin{equation}\label{t-Q-ss}
\ds \tp(\lambda\,;\state)\;
\underset{\state\mapsto\state'}{\Q_\ff} \;=\;
\underset{\state\mapsto\state'}{\Q_\ff} \tp(\lambda\,;\state')\;.
\end{equation}

\subsection{Solutions of the cyclic boundary conditions}

Careful and rather tedious investigations of the quantum Darboux
relations (\ref{LL-darboux},\ref{op-recursion}) and cyclic
boundary conditions (\ref{cyclic-recursion}) allows us to
formulate the following
\begin{prop}\label{prop-recursion}
Let the state $\state$ is $n$-solitonic one  with the data
$\{f_k\}_{k=1}^n$. Then all possible solutions of
(\ref{cyclic-recursion}) gives the following mappings $\state$
$\mapsto$ $\state'$, classified by the values of the spectral
parameter $\phi$ of $\Q_\ff$:
\begin{itemize}
\item[a).] if $\phi\not\in\,\{\phi_k\}_{k=1}^n$, but
$\phi$ belongs to set (\ref{phi-k}), then $\state'$ is
$n+1$-solitonic state with the data
$\{f_k',\phi_k\}_{k=1}^{n+1})$, where
\begin{equation}
\ds f_k'\;=\;f_k\,s(\phi_k,\phi)\;,\;\;\;k=1..n\;,
\end{equation}
$s$ is given by (\ref{s-function}), and
\begin{equation}
\phi_{n+1}\;=\;\phi\;,\;\;\;\;f_{n+1}'\,-\,\mbox{arbitrary
complex number}\;.
\end{equation}
\item[b).] If the cyclicity condition is not satisfied,
i.e. $\EXP^{2\,i\,M\,\phi}\neq 1$, then the previous formulae are
valid, but 
with $f'_{n+1}\equiv 0$.
\item[c).] If $\phi=-\phi_n$, then $\state'$
is the $n-1$ solitonic state with the data
$\{f_k\,s(\phi_k,\phi),\phi_k\}_{k=1}^{n-1})$.
There is an annihilation of the soliton due to zero of the function
$s(\phi_n,-\phi_n)$.
\item[d).] If $\phi=\phi_n$, then the state $\state'$ corresponds to
$n-1$ solitonic state with the data $\{f_k s(\phi_k,\phi)
d(\phi_k,\phi),\phi_k\}_{k=1}^{n-1})$ up to a gauge multipliers.
\end{itemize}
\end{prop}
As it should be, (\ref{LL-darboux}) defines actually the action
of the vertex operator of the B\"acklund transformation, creating
or annihilating a soliton. Note also, not all the four items of
proposition (\ref{prop-recursion}) are useful. Namely, we may
ignore item `d)': it is equivalent to the item `c)'. One may
show, `d)' and `c)' produces the same $U_m'$ and $W_m'$ due to
$\Lambda(\phi)=\Lambda(-\phi)$ and $\Om^*(\phi)=\Om(-\phi)$.

\subsection{Permutation of $\Q_\ff$-operators}

Consider now the scenario `a)' of the proposition -- the creation
of the new soliton. Besides the quantization of $\phi$ this means
also the additional degree of freedom in the final state: an
arbitrary amplitude $f$ of the $\phi$-th soliton. Following is the
visual notation for it:
\begin{equation}
\ds\state\;\xrightarrow{f,\phi}\;\state'\;.
\end{equation}
It is useful to exhibit the parametric dependence on $f$:
$\ds\underset{\state\mapsto\state'}{\Q_\ff}(f)$.

Compare now two successive creations of solitons: the first one
\begin{equation}
\ds\state\; \xrightarrow{f,\phi}\; \state'\;
\xrightarrow{f',\psi}\; \state''\;,
\end{equation}
and the second one
\begin{equation}
\ds\state\; \xrightarrow{g,\psi}\; \widetilde{\state}'\;
\xrightarrow{g',\phi}\; \state''\;.
\end{equation}
The initial and the final states are the same, this provides
\begin{equation}
\ds g'\;=\;f\,s(\phi,\psi)\;,\;\;\; f'\;=\;g\,s(\psi,\phi)\;.
\end{equation}
It follows the equivalence:
\begin{equation}\label{QQ-QQ}
\ds \underset{\state\mapsto\state'}{\Q_{\phi}}(f) \,\cdot\,
\underset{\state'\mapsto\state''}{\Q_{\psi}} (g\,s(\psi,\phi))
\;\sim\; \underset{\state\mapsto
\widetilde{\state}'}{\Q_{\psi}}(g) \,\cdot\,
\underset{\widetilde{\state}'\mapsto\state''}{\Q_{\phi}}
(f\,s(\phi,\psi))\;.
\end{equation}
The sign `$\sim$' means the existence of some extra multipliers
in such pseudocommutation relations.

Consider now the successive creation of the maximal solitonic
state. Let
\begin{equation}
\ds g_1=f_1\;,\;\;\;g_k=f_k\prod_{j<k}\,s(\phi_k,\phi_j)\;.
\end{equation}
Consider
\begin{equation}\label{S-matrix}
\ds\smatrix(\{f_k\}_{k=1}^{M-1})\;\stackrel{def}{=}\;
\underset{\state_0\mapsto\state_1}{\Q_{\phi_1}}(g_1)
\underset{\state_1\mapsto\state_2}{\Q_{\phi_2}}(g_2) \cdots
\underset{\state_{M-2}\mapsto\state_{M-1}}
{\Q_{\phi_{M-1}}}(g_{M-1})\;,
\end{equation}
where the initial state $\state_0$ is supposed to be the
homogeneous one. Up to multipliers, operator-valued function
$\smatrix$ is symmetrical with respect to any permutation of
$\{f_k\}_{k=1}^{M-1}$, this follows from (\ref{QQ-QQ}). The final
state $\state_{M-1}$ of $\smatrix$ has the solitonic data
$\{F_k\}_{k=1}^{M-1}$, where
\begin{equation}
\ds F_k\;=\;f_k\,\prod_{j\neq k}\,s(\phi_k,\phi_j)\;.
\end{equation}
In general all $\Q_\ff$ are invertible, and so invertible
$\smatrix$ obeys
\begin{equation}
\ds
\tp(\lambda;\state_0)\;\smatrix\;=\;\smatrix\;\tp(\lambda;\state_{M-1})\;,
\end{equation}
this 
proves proposition (\ref{prop-isospectral}) for the given initial
and final states.

\subsection{Baxter's equation for operator $\Q_\ff$}

Auxiliary quantum Lax operator (\ref{ldst-uwkappa}) factorizes
when $\lambda=\lambda_\ff$:
\begin{equation}\label{degen}
\ds \ldst_{\ff}(\lambda_\ff)\;=\;
{1-\omeg\kappa_\ff^{}\wop_\ff^{}
\choose
-\omeg\lambda_\ff^{}\uop_\ff^{-1}\wop_\ff^{}}\;\cdot\;
\biggl(\;1\;,\;-\omeg\lambda_\ff^{-1}\uop_\ff^{}\biggr)\;.
\end{equation}
This degeneration leads to the Baxter equation in the operator
form by rather usual procedure of the triangulization of the
monodromy matrix \cite{Baxter,BS}. Details are inessential here,
and the final answer reads
\begin{equation}\label{BE-operator-12}
\ds \tp(\lambda_\ff\,;\state)\;\Q_\ff\;=\;
\Q_\ff\;\tp(\lambda_\ff\,;\state')\;=\;
\Q^{(1)}_\ff\;+\;\Q^{(2)}_\ff\;,
\end{equation}
where in the operator form
\begin{equation}\label{Q12}
\ds \Q_\ff^{(1)}\;=\;{1\over\mu_\ff}\,\Xop\,\Q_\ff\,\Xop^{-1}\;,
\;\;\;
\Q_\ff^{(2)}\;=\;{\mu_\ff\over\lambda_\ff^M}\,\Zop\,\Q_\ff\,\Xop\;.
\end{equation}
$\Xop$ and $\Zop$ are given by (\ref{XYZ}). Multiplier $\mu_\ff$
is originated from the detailed consideration of the recursion
(\ref{op-recursion}) and (\ref{cyclic-recursion}), providing
\begin{equation}
\ds (\lambda_\ff,\mu_\ff)\;\in\;\Gamma_M\;.
\end{equation}
In the special case when $\state=\state'=\state_0$ (this is the
case `b' of proposition (\ref{prop-recursion})), $\lambda_\ff$ is
arbitrary complex number, and eq. (\ref{BE-operator-12}) becomes
usual Baxter's relation.

\subsection{Meaning of the modified $\Q$-operators}

Consider now (\ref{ltQ}) in the degeneration point (\ref{degen})
for the whole chain. It is easy to rewrite (\ref{ltQ}) in the form
\begin{equation}\label{t-mon-1}
\ds\begin{array}{l} \ds\left(\mathbf{a}(\lambda_\ff)\,-\,
\omeg\,{u_{\ff,1}\over\lambda_\ff}\,
\xop_\ff^{}\mathbf{c}(\lambda_\ff)\right)
\,\cdot\,\widehat{\Q}_\ff \,=\, \widehat{\Q}_{\ff}^{(1)}\;,\\
\\
\ds\left(\mathbf{b}(\lambda_\ff)\,-\,
\omeg\,{u_{\ff,1}\over\lambda_\ff}\,
\xop_\ff^{}\mathbf{d}(\lambda_\ff)\right)
\,\cdot\,\widehat{\Q}_\ff \,=\,
-\omeg\,{u_{\ff,M+1}\over\lambda_\ff}\,\widehat{\Q}_{\ff}^{(1)}\,
\xop_\ff^{}\;,
\end{array}\end{equation}
or, in the equivalent form
\begin{equation}\label{t-mon-2}
\ds\begin{array}{l} \ds
\mathbf{a}(\lambda_\ff)\,\widehat{\Q}_\ff\,
\omeg\,{u_{\ff,M+1}\over\lambda_\ff}\,\xop_\ff^{}\,+\,
\mathbf{b}(\lambda_\ff)\,\widehat{\Q}_\ff \,=\,
\omeg\,{u_{\ff,1}\over\lambda_\ff}\,\xop_\ff^{}\,
\widehat{\Q}_{\ff}^{(2)}\;,\\
\\
\ds \mathbf{c}(\lambda_\ff)\,\widehat{\Q}_\ff\,
\omeg\,{u_{\ff,M+1}\over\lambda_\ff}\,\xop_\ff^{}\,+\,
\mathbf{d}(\lambda_\ff)\,\widehat{\Q}_\ff \,=\,
\widehat{\Q}_{\ff}^{(2)}\;,\\
\end{array}\end{equation}
It is interesting to consider these relations for the mapping
$\state_0\mapsto\state_1$, where $\state_0$ is the homogeneous
case, and $\state_1$ is one-solitonic state with the wave number
$\phi$ and the amplitude $f$. Note, $\mu_\ff$ and $\lambda_\ff$
do not depend on $f$. Consider now the limit
\begin{equation}
\ds f\;\mapsto\;-1\;.
\end{equation}
In this limit state $\state_1$ has the remarkable feature:
$u_{\ff,1}\;\equiv\;u'_1\;=\;0$ . Then from (\ref{t-mon-1}) and
(\ref{t-mon-2}) it follows:
\begin{equation}\label{bQ=0}
\ds \mathbf{b}(\lambda_\ff)\cdot
\underset{\state_0\mapsto\state_1}{\Q_{\ff}}\;=\;0\;.
\end{equation}
Due to the symmetry of $\smatrix(\{f_k\}_{k=1}^{M-1})$ with
respect to any permutation of $\{f_k\}$, one may conclude
\begin{equation}\label{main-relations}
\ds \mathbf{b}(\lambda_{\phi_n})\,\cdot\,
\smatrix(\{f_k=-1\}_{k=1}^{M-1})\;=\;0\;\;\;
\forall\;\;n\;=\;1,...,M-1\;.
\end{equation}
This means that in the special point $f_k=-1$ operator $\smatrix$
becomes rather degenerative, and in the $N$-dimensional projector
decomposition of $\smatrix$ all left vectors may be regarded as
the eigenvectors of operator $\mathbf{b}(\lambda)$ with zeros
$\lambda_{\phi_k}$, see eq. (\ref{lambda-phi-k}) and the
discussion around it.

\section{Discussion}

The main object, appeared in this note, is the operator
$\smatrix(\{f_k=-1\}_{k=1}^{M-1})$. Recall once more, we have the
explicit form of the matrix elements of each $\Q_\ff$-operator,
entering to $\smatrix$. Note, there are some arbitrariness in
their definition. The technical reason of it is that one has to
deal with the ambiguities $0/0$, and the principal reason is that
the right vectors of $\smatrix$ may be chosen in rather arbitrary
way. There still exists the problem of the most useful choice of
these right vectors. All the details concerning the explicit form
of $\Q_\ff$ and $\smatrix$, as well as the exact form of the
pseudocommutation relation (\ref{QQ-QQ}) will be the subjects of
the subsequent papers. The other subject to be mentioned is the
strict investigation of the classical counterpart in the
application to the quantum technique. The final aim is the exact
solution of the model, i.e. the solution of Baxter's equation an
the construction of the eigenvectors of $\tp(\lambda)$, at least
with the help of the separation of the variables method.

What is to be mentioned else. The ``dimer self-trapping'' Lax
matrix at the root of unity belongs to the class of the chiral
Potts model Lax matrices \cite{BS}. The quantum intertwiner of
DST Lax matrices is a particular case of the chiral Potts model
$S$-matrix. The method of applying a nontrivial classical
dynamics to the chiral Potts (or DST - it is much more simple)
model will give the analogous scheme for the separation of the
variables method, including the exact construction for the
eigenvectors of the elements of the CPM monodromy matrix. These
we will do separately.

{\bf Acknowledgements} The authors are grateful to R. Baxter, V.
Bazhanov, V. Mangazeev, G. Pronko, E. Sklyanin, A. Belavin, Yu.
Stroganov and A. Isaev for useful discussions and comments.

\bibliographystyle{amsplain}

\begin{thebibliography}{**}


\bibitem{PS}
G. Pronko and S. Sergeev, ``Relativistic Toda Chain'', {\em
nlin.SI/0009027}, to appear in {\em Journal of Applied
Mathematics}.

\bibitem{CPM}
R.J. Baxter, J.H.H. Perk and H. Au-Yang, ``New solutions of the
star-triangle relations for the chiral Potts models'', {\em Phys.
Lett.} {\bf  A128} (1988) 138-142.


\bibitem{BS}
V. V. Bazhanov and Yu. G. Stroganov, ``Chiral Potts model as a
descendant of the six-vertex model'', {\em J. Stat. Phys.} {\bf
59} (1990) 799-817.

\bibitem{Sklyanin-sep}
E. K. Sklyanin, {\em Lecture Notes in Physics}, {\bf 226} (1985)
196;

\bibitem{Sklyanin-rev}
E. K. Sklyanin, ``Baecklund transformations and Baxter's Q -
operator'', {\em preprint nlin.SI/0009009}.

\bibitem{KS-manybody}
V. B. Kuznetsov and E. K. Sklyanin, ``On Baecklund
transformations for many-body systems'', {\em J. Phys. A: Math.
Gen.} {\bf 31} (1998) 2241-2251.


\bibitem{Lebedev}
S. Kharchev, and D. Lebedev, ``Integral representation for the
eigenfunctions of a quantum periodic Toda chain'', {\em Letters in
mathematical physics}, {\bf 50} (1999) 53-77.

\bibitem{BR-unpublished}
V. V. Bazhanov and N. Yu. Reshetikhin, ``Chiral Potts model and
discrete Sine-Gordon model at roots of unity'', unpublished
(1995).

\bibitem{Sergeev}
S. Sergeev, ``Quantum 2+1 evolution model'', {\em J. Phys. A:
Math. Gen.} {\bf 32} (1999) 5693-5714; ``Solitons in a 3d
integrable model'', {\em Phys. Lett.} {\bf A 265} (2000) 364-368;
``On exact solution of a Classical 3d integrable model'', {\em
Journal of Nonlinear Mathematical Physics} Vol 7 No 1 (2000)
57-72; ``Auxiliary transfer matrices for three dimensional
integrable models'', {\em Theoretical and Mathematical Physics},
Vol 124, No 3 (2000) 1187-1201. ``Some properties of local linear
problem's matrices'', in: {\em Izergin's memorial volume,
proceedengs of S-PBMI}, in russian.

\bibitem{Baxter}
R. J. Baxter, ``Exactly Solved Models in Statistical Mechanics'',
Academic Press, London, 1972.


\end{thebibliography}

\end{document}